# Dissociation of ssDNA – Single-Walled Carbon Nanotube Hybrids by Watson-Crick Base Pairing

Seungwon Jung[1], Misun Cha[2*], Jiyong Park[3], Namjo Jeong[1, 4], Gunn Kim[3], Changwon Park[3], Jisoon Ihm[3], and Junghoon Lee[1, 2*]

[1]*School of Mechanical and Aerospace Engineering, Seoul National University, Seoul 151-742, Korea*

[2]*Institute of Advanced Machinery and Design, Seoul National University, Seoul 151-744, Korea*

[3]*FPRD and Department of Physics and Astronomy, Seoul National University, Seoul 151-742, Korea*

[4]*Nanomaterials Research Center, Korea Institute of Energy Research, Daejeon 305-343, Korea*

Corresponding authors: jleenano@snu.ac.kr, cmsbest@snu.ac.kr

Single strand DNA (ssDNA) forms a hybrid with a single walled carbon nanotube (SWNT) through π-stacking between the bases of the ssDNA and the atomic structures on the sidewall of the SWNT[1, 2]. The ssDNA-SWNT hybrid has gained interests as a new nano-bio conjugate for various applications such as drug delivery[3], bio/chemical sensors[4, 5, 6], and the sorting of SWNTs[7, 8]. It has also been revealed that the optical and the electrical properties of the SWNT may be altered by the interaction between the ssDNA and the SWNT[9-11].

A next exploration of significance would be regarding the aftermath of a Watson-Crick base paring or hybridization by introducing a DNA sequence complementary to the ssDNA attached on the SWNT. The hybridization has been confirmed by electrochemical and optical methods such as the shift of field effect transistor (FET) characteristics[5] and band-gap fluorescence modulation[6] while the reaction with non-complementary DNA resulted in no change as predicted[5, 6, 12]. However, the detailed mechanism and the consequence of the hybridization are still in controversy. Some papers suggest that the double strand DNA (dsDNA) will remain attached along the aromatic group on the sidewall of the SWNT after the base pairing[5, 6]. Others, on the other hand, favor the dissociation from the sidewall of the SWNT as the dsDNA is formed[12, 13].

In this paper, we report experimental and theoretical proofs that unambiguously support the dissociation of the hybridized dsDNA from the SWNT. This wrapping-unwrapping transition is primarily evidenced by the shift of electrical properties, monitored through an FET-type measurement. As previously reported, a metallic SWNT, when forming a hybrid with an ssDNA, shows a p-type semiconducting behavior[11]. Here we demonstrate that the electrical property of hybridized product essentially returns to the metallic state caused by the dissociation event. This unwrapping process is also confirmed through gel electrophoresis, and further verified with the Raman spectroscopy. We also used molecular dynamics (MD) simulations and binding energy analyses to study detailed mechanisms.

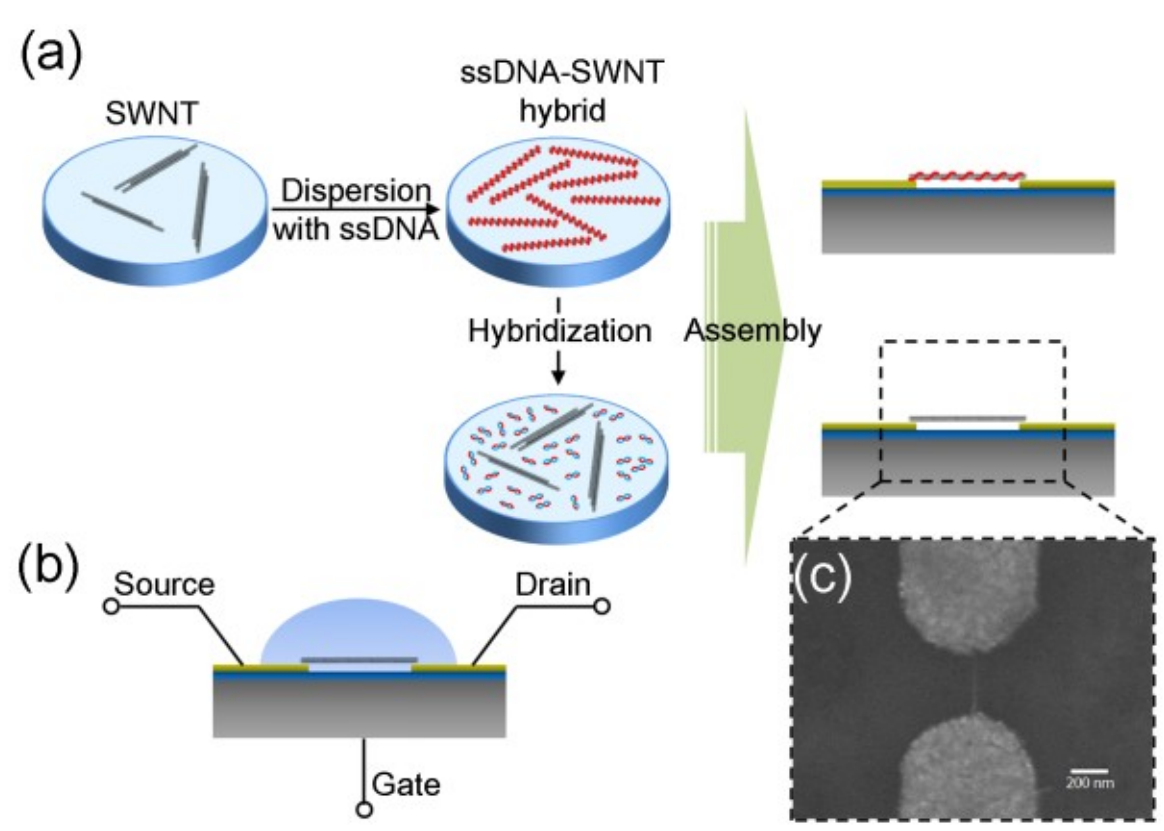

**Figure 1**. Schematic of monitoring DNA hybridization. (a) The procedure of DNA hybridization experiment. (b) Schematic of the FET measurement. (c) Scanning electron microscopy (SEM) image of the assembled SWNT

Figure 1 shows the schematic of the experimental procedure and the FET measurement. Initially 0.1 mg of the SWNTs synthesized through the high-pressure carbon monoxide conversion process (HiPco, Carbon Nanotechnologies, Inc., TX, USA) were dispersed with the 10 μM ssDNAs by using sonication and centrifugation in de-ionized (DI) water[11]. Since there was excessive amount of SWNTs, the ssDNAs were completely consumed by forming the hybrids that were well dispersed in water. The unbound SWNTs were removed by centrifugation. Our observation with a spectrophotometer (NANODROP 2000, Thermo Fisher Scientific Inc., USA) indicates that the ssDNAs were completely consumed during the formation of the ssDNA-SWNT hybrids (see the Supporting Information). The resulting solution with the ssDNA-SWNT hybrids was diluted by 10 times with DI water, followed by the hybridization process in which 1 μM complementary DNAs (cDNAs) were injected into the ssDNA-SWNT hybrid solution. The hybridization was carried out in a buffer solution (50 mM PBS buffer, equivalently mixed during the hybridization) to improve the efficiency[14].

The sample after the reaction was dielectrophoretically (at 5 MHz & 5 V for 1 min) deposited between the electrodes with 300 nm gap in 30 pairs which were pre-fabricated through nanoimprint lithography (NIL) [11]. The dielectrophoretic alignment and assembly under the given conditions resulted in predominantly metallic SWNTs captured and deposited even when the SWNTs were wrapped with DNAs[15, 16]. The FET measurement was used to obtain a source-drain current according to the sweeping of gate voltage from -15 V to +15 V at a constant source-drain bias voltage of 100 mV[17].

Figure 2 shows the electrical properties of the ssDNA-SWNT hybrids measured before and after the hybridization. The hybridization experiment of Figure 2a used a homogeneous sequence: 18-mer poly(C)-SWNT hybrid and poly(G) target with the same length. The signal was normalized by the maximum value of the measured current (raw graphs and hysteresis are included in Supporting Information). It has been previously reported that the electronic structure of the metallic SWNT was significantly affected by the helical wrapping of the ssDNA, resulting in the semiconducting behavior in Figure 2[11]. This semiconducting property of the ssDNA-decorated metallic SWNT was caused by the electron transfer from the SWNT to the ssDNA. Water molecules were found critical to activate this metal-semiconductor transition in the ssDNA-SWNT hybrid. Thus every reaction and measurement in this paper were carried out in wet state. As a result of the hybridization with cDNA, the current response of the SWNT was shifted upward and became almost flat, reclaiming the characteristic of the original metallic behavior. A similar outcome was observed when a heterogeneous sequence was used (Figure 2b). The sequence used for this case (target DNA: 5'-ccg acc gac gtc ggt tgc-3') has been selected for the detection of *human papillomavirus* (HPV)[18]. These results show that the shift of the conducting property resulting from the interaction of DNA is essentially sequence-independent. On the contrary, such a shift was not observed in case of the reaction with mismatching DNA. This experimental result describes no interaction of the hybrid with mismatching sequence, which is consistent with previous studies[5, 6, 12].

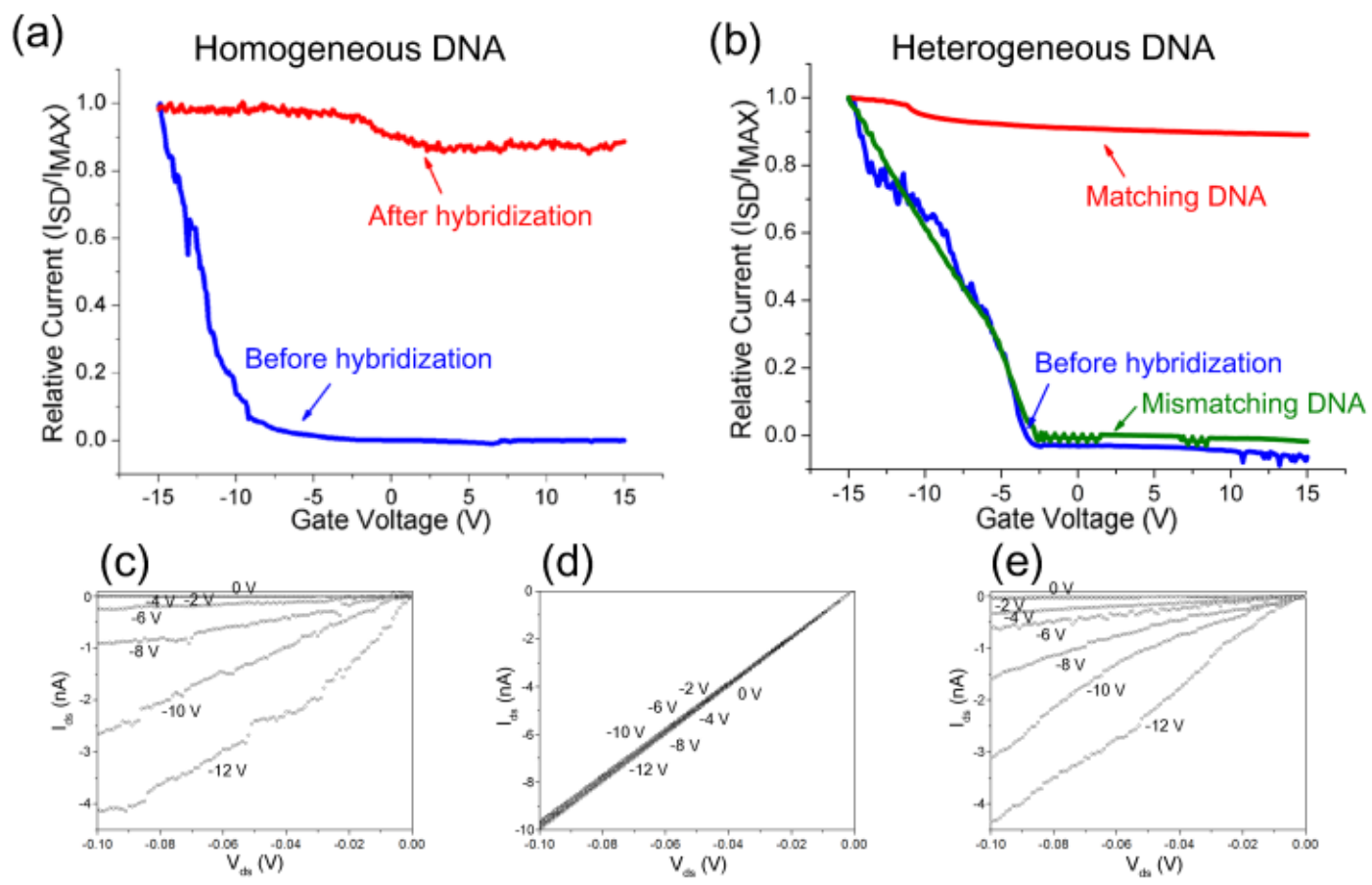

**Figure 2**. Results of hybridization between the hybrid and the complementary DNA with homogeneous (a) and heterogeneous (b) sequences. (c), (d), and (e) show Ids-Vds data for several gate voltage in case of hybrid, hybridization with matching cDNA and mismatching DNA respectively.

In figure 3a we report a gel electrophoresis (15% polyacrylamide gel) reveals the fate of the dsDNA released as a result of the hybridization. Fluorescence signal did not appear on the lane #1 loaded with ssDNA-SWNT because hybrids were trapped into the well owing to gel condition of high concentration. The Ethidium Bromide (EtBr) dye embedded in the polyacrylamide gel does not stain the ssDNA as effectively as the dsDNA, leading to virtually weak fluorescence signal in the lane #2. The lane #3 was loaded with the pure hybridized product of the ssDNA and its cDNA as a control, resulting in the strong fluorescence. The lane #4 and #5 were loaded with the mixture of the ssDNA-SWNT hybrid and the complementary and mismatching DNA sequence respectively. The fluorescence location in the lane #4 coincides with that of the control in the lane #3. However, the fluorescence signal in the lane #5 was observed in the same location with ssDNA (lane #2). Furthermore, fluorescence band of residual ssDNA was observed in the lanes #3 and #4. These results indicate that dsDNA originated from the ssDNA released from the SWNT during the hybridization.

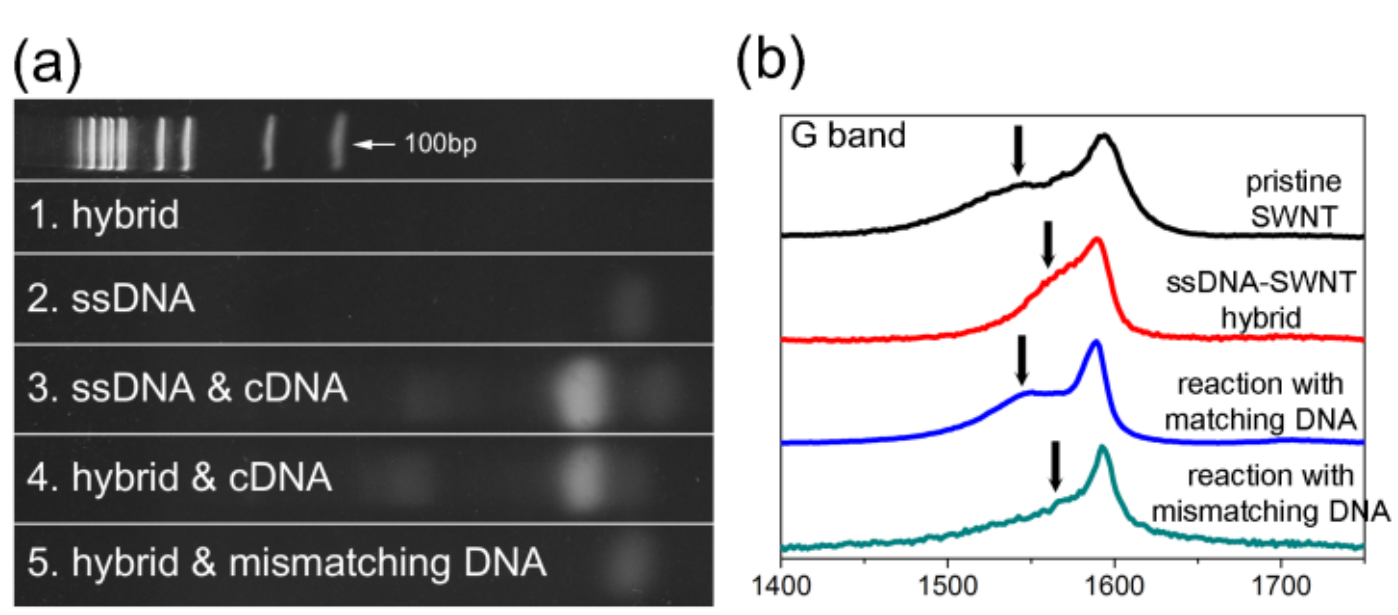

**Figure 3**. Experimental evidences for unwrapping due to hybridization with (a) the gel electrophoresis (b) tangential bands (G-bands) in Raman spectra.

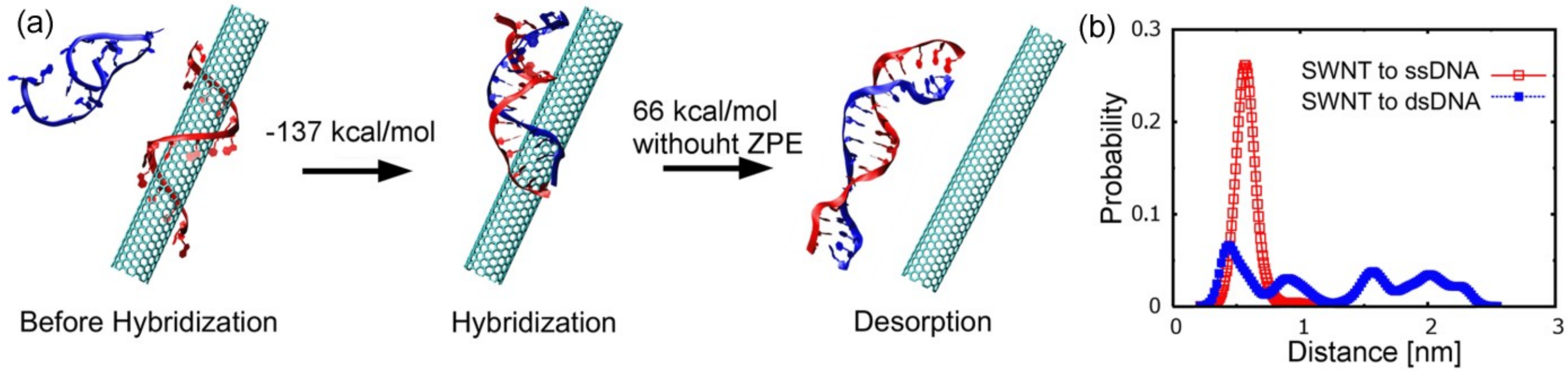

**Figure 4**. Energy change and the detachment of dsDNA associated with the hybridization reaction. (a) Binding energy differences between relatively stable states. (b) The distance from the SWNT surface to phosphate atoms in DNA molecules adsorbed on the SWNT surface was measured.

The Raman spectroscopy was used to further investigate the unwrapping event due to the hybridization. Radial breathing modes (RBMs) and tangential modes (G-bands) were measured for the pristine SWNT and the ssDNA-SWNT hybrids in water before and after the hybridization. A micro-Raman system (LabRam HR, Jobin-Yvon, France) was used with laser lines at a wavelength of 514.5 nm from an Ar-ion laser. The diameter of the SWNT used in this study was determined by the analysis of the RBM peaks[19] (RBM peaks are described in Supporting Information). It was 1.1 $\pm$ 0.2 nm for all three cases. The G-band in Figure 3 shows a noticeable difference between the states before and after the hybridization. The pristine SWNT shows a broad and asymmetric Breit-Wigner-Fano (BWF) line shape (vertical arrow in Figure 3) unique to the Raman spectra of the metallic SWNT[20]. The BWF line of the ssDNA-SWNT hybrids in solution was shifted to a higher frequency, which agrees with previous results[11]. After the hybridization with the cDNA, however, the BWF line clearly shifted back to a lower frequency that corresponds to the metallic property regained by unwrapping.

To understand the atomistic mechanism of the ssDNA-SWNT hybrid formation and the DNA hybridization reaction observed in our experiment, we carried out MD simulations and subsequent binding energy analyses. We identified several states of interest whose energy differences drive overall progress of the chemical reactions observed in our experiments (Figure 4a). The binding energy was approximated to be the sum of vacuum, polar, and nonpolar solvation free energies as well as trans-rotational and vibrational entropic contributions[21]. A copy of ssDNA (5'-ccg acc gac gtc ggt cg-3') initially wrapped around an (8,8) armchair type SWNT and its cDNA (5'-cga ccg acg tcg gtc gg-3') was solvated in medium. The nucleic acid base of ssDNA made a tight contact with the SWNT; the average distance from the SWNT to nucleic acid base atoms of DNA was 0.43 nm and the distribution of the distance was sharply peaked around its mean (Figure 4b). The backbone phosphate atom of ssDNA was 0.15 nm away from SWNT relative to the base atoms, which reclaimed the adsorption of ssDNA was mediated by the base to SWNT interaction. These findings were in good agreement with our experimental result. As the two ssDNAs constructed a dsDNA on the SWNT surface, energy (134 kcal/mol) was released (an exothermic process).

After the hybridization, the dsDNA could be detached from the SWNT because of the weaker binding energy. The surface groove of the dsDNA may still make contacts with the SWNT[22]. However, the contact became less intact relative to the ssDNA-SWNT complex. Before the hybridization, the number of atomic contacts between the ssDNA and the SWNT was 527, while it became 114 after the hybridization. The distance from the SWNT to the phosphate atoms of the dsDNA was 1.25 nm and the distribution became dispersed after the hybridization reaction (Figure 4b). In addition, dsDNA diffused more freely on the SWNT surface than the ssDNA; root-mean-squared-deviation (RMSD) of the ssDNA along the SWNT axis throughout the last 5 ns MD simulation trajectory was 5.87 Å, whereas it was 8.10 Å for the case of dsDNA on the SWNT after the hybridization. On the other hand, 66 kcal/mol was needed to detach the dsDNA from the SWNT surface according to the method described at the end (without zero-point energy correction). If we should include the zero-point energy corrections, this number would be reduced to ~40 kcal/mol. In addition, the hybridized dsDNA would be vulnerable to external perturbations such as agitation and heating (which could happen during the hybridization process in general) and could easily be detached from the SWNT. Based on these observations, we conclude that spontaneous desorption of some dsDNAs from the SWNT surface was likely to occur following the Watson-Crick base paring between ssDNAs. To recapitulate, the hybridization and the subsequent detachment of the dsDNA were key ingredients for the recovery to the metallic characteristics, as they could substantially reduce the electron transfer from the adsorbed ssDNA to the SWNT.

In summary, the unwrapping event of ssDNA from the SWNT during the Watson-Crick base paring was confirmed through electrical and optical methods, and binding energy calculations. While the ssDNA-metallic SWNT hybrid showed the p-type semiconducting property, the hybridization product recovered metallic properties. The gel electrophoresis directly verified the result of wrapping and unwrapping events which was also reflected to the Raman shifts. The MD simulations and binding energy calculations provided atomistic description for the pathway to this phenomenon. This nano-physical phenomenon will open up a new approach for nano-bio sensing of specific sequences with the advantages of efficient particle-based recognition, no labeling, and direct electrical detection which can be easily realized into a microfluidic chip format.

**Acknowledgement.** We thank M. K. Choi in Raman Research Center for fruitful discussion. This work was supported by the Pioneer R&D Program for Converging Technology and Basic Research Promotion Fund (Grant Number M10711270001-08M1127-00110), the National Research Foundation of Korea (NRF) (R0A-2007-000-10051-0) (M.C. & J.L.), the second BK21 program (G.K.), and the Center for Nanotubes and Nanostructured Composites (C.P. & J.I.), all through the Korea Science and Engineering Foundation (KOSEF) funded by the Ministry of Education, Science and Technology (MEST). This work was also supported by the Samsung Advanced Institute of Technology (SAIT, S.J. & J.L.). Fabrication and experiments were performed at the Inter-university Semiconductor Research Center (ISRC) in Seoul National University. The computations were performed through the support of the Korean Institute of Science and Technology Information (KISTI).

.

REFERENCES

(1) Zheng, M.; Jagota, A.; Semke, E. D.; Diner, B. A.; Mclean R. S.; Lustig, S. R.; Richardson, R. E.; Tassi, N. G. *Nat. Mat.* **2003**, *2*, 338-342.

(2) Zheng, M.; Jagota, A.; Strano, M. S.; Santos, A. P.; Barone, P.; Chou, S. G.; Diner, B. A.; Dresselhaus, M. S.; Mclean R. S.; Onoa, G. B.; Samsonidze, G. G.; Semke, E. D.; Usrey, M.; Walls, D. J. *Science* **2003**, *302*, 1545-1548.

(3) Kam, N. W. S.; O'Connell, M.; Wisdom, J. A.; Dai, H. *Proc. Natl. Acad. Sci. USA* **2005**, *102*, 11600-11605.

(4) Staii, C.; Chen, M.; Gelperin, A.; Johnson, A. T. Jr. *Nano Lett.* **2005**, *5(9)*, 1774-1778.

(5) Star, A.; Tu, E.; Niemann, J.; Gabriel, J.-C. P.; Joiner, S.; Valcke, C. *Proc. Natl. Acad. Sci. USA* **2006**, *103*, 921-926.

(6) Jeng, E. S.; Moll, A. E.; Roy, A. C.; Gastala, J. B.; Strano, M. S. *Nano Lett.* **2006**, *6(3)*, 371-375.

(7) Tu, X.; Manohar, S.; Jagota, A.; Zheng, M. *Nature* **2009**, *460*, 250–253.

(8) Hersam, M. C. *Nat. Nanotechnol.* **2008**, *3*, 387–394.

(9) Heller, D. A.; Jeng, E. S.; Yeung, T.-K.; Martinez, B. M.; Moll, A. E.; Gastala, J. B.; Strano, M. S. *Science* **2006**, *311*, 508-511.

(10) Kawamoto, H.; Uchida, T.; Kojima, K.; Tachibana, M. *J. Appl. Phys.* **2006**, *99*, 094309.

(11) Cha, M.; Jung, S.; Cha, M.-H.; Kim, G.; Ihm, J.; Lee, J. *Nano Lett*. **2009**, *9(4)*, 1345-1349.

(12) Chen, R. J.; Zhang, Y. *J. Phys. Chem. B* **2006**, *110(1)*, 54-57.

(13) Karachevtsev, V. A.; Gladchenko, G. O.; Karachevtsev, M. V.; Glamazda, A. Y.; Leontiev, V. S.; Lytvyn, O. S.; Dettlaff-Weglikowska, U. *Mol. Cryst. Liq. Cryst.* **2008**, *497*, 339-351.

(14) Sambrooke, J.; Russel, D. W. *Molecular Cloning: A Laboratory Manual,* 3rd ed.; Cold Spring Harbor Laboratory Press; New York, 2001; ch. 10.

(15) Krupke, R.; Hennrich, F.; Löhneysen, H.; Kappes, M. M. *Science* **2003**, *301*, 344-347.

(16) Sickert, D.; Taegar, S.; Neumann, A.; Jost, O.; Eckstein, G.; Mertig, M.; Pompe, W. *AIP Conf. Proc.* **2005**, *786*, 271-274.

(17) Kim, W.; Javey, A.; Vermesh, O.; Wang, Q.; Li, Y.; Dai, H. *Nano Lett.* **2003**, 3 ( 2) 193–198

(18) Rashmi, S. H.; Steven, R. G.; Laimonis, A. L.; Paul, B. S. *Nature* **1992**, *359*, 505 - 512.

(19) Saito, R.; Takeya, T.; Kimura, T.; Dresselhaus, G.; Dresselhaus, M. S. *Phys. Rev. B* **1998**, *57(7)*, 4145-4153.

(20) Brown, S. D. M.; Jorio, A.; Corio, P.; Dresselhaus, M. S.; Dresselhaus, G.; Saito, R.; Kneipp, K. *Phys. Rev. B* **2001**, *63*, 155414-155421.

(21) Wang, W.; Kollman, P. A. *J. Mol. Biol.* **2000**, *303*, 567-582.

(22) Zhao, X.; Johnson, J. K. *J. Am. Chem. Soc.* **2007**, *129*, 10438-10445.

(23) Chi, Q.; Göpel, W.; Ruzgas, T.; Gorton, L.; Heiduschka, P. *Electroanalysis* **2005**, *9(5)*, 357-365.